\documentstyle[preprint,floats,pra,aps]{revtex}
\tightenlines
\begin{document} \draft

\title{\Large \bf Interferometers and Decoherence Matrices}

\author{D. Han\footnote{electronic mail: han@trmm.gsfc.nasa.gov}}
\address{National Aeronautics and Space Administration, Goddard Space
Flight Center, Code 935, Greenbelt, Maryland 20771}

\author{Y. S. Kim\footnote{electronic mail: yskim@physics.umd.edu}}
\address{Department of Physics, University of Maryland, College Park,
Maryland 20742}

\author{Marilyn E. Noz \footnote{electronic mail: noz@nucmed.med.nyu.edu}}
\address{Department of Radiology, New York University, New York, New York
10016}

\maketitle

\begin{abstract}
It is shown that the Lorentz group is the natural language for two-beam
interferometers if there are no decoherence effects.  This aspect of
the interferometer can be translated into six-parameter representations
of the Lorentz group, as in the case of polarization optics where there
are two orthogonal components of one light beam.  It is shown that there
are groups of transformations which leave the coherency or density
matrix invariant, and this symmetry property is formulated within
the framework of Wigner's little groups.  An additional mathematical
apparatus is needed for the transition from a pure state to an impure
state.  Decoherence matrices are constructed for this process, and their
properties are studied in detail.  Experimental tests of this symmetry
property are possible.

\end{abstract}


\narrowtext

\section{Introduction}\label{intro}

In our earlier papers~\cite{hkn97josa,hkn97,hkn99,barakat63}, we have
formulated polarization optics in terms
of the two-by-two and four-by-four representations of the six-parameter
Lorentz group.  It was noted that the two-component Jones vector and
the four-component Stokes parameters are like the relativistic spinor
and the Minkowskian four-vector respectively.  We were able to
identify the attenuator, rotator, and phase shifter with appropriate
transformation matrices of the Lorentz group.  It was noted that
the two-element Jones vector is like the two-component Pauli spinor
and that the four Stokes parameters act like the elements of a
Minkowskian four-vector.

The purpose of this paper is to show that the mathematics of
polarization optics is applicable also to interferometers.  Our
reasoning is that polarization optics is basically the physics
of two plane waves.  The same is true for two-beam interferometers.
We need mathematical devices which
will perform phase shifts between the waves and which will take care of
attenuations at different rates.  In the case of interferometers, it
is possible to achieve the beam split and synthesis by rotation matrices.
We can use the matrices of the above-mentioned Lorentz group in order
to achieve these basic physical operations.

In addition, in this paper, we discuss the mathematical device which
will describe the decoherence effect due to random phases.  For this
purpose, we need density matrices.  However, the coherency matrix
serves as the density matrix, and its four elements constitute the four
components of the Stokes vector~\cite{born80,fey72}.  It was noted in
our previous paper that it is possible to construct a four-by-four
decoherence matrix which will transform a pure-state Stokes vector
into a mixed-state Stokes vector.  Unlike the case of attenuations,
rotations, or beam splits and syntheses, the decoherence matrix does
not belong to the Lorentz group.

In order to study the decoherence process more carefully, we borrow
the concept of Wigner's little group originally developed for
studying internal space-time symmetries of elementary
particles~\cite{wig39,knp86}.  Wigner's little group is the maximal
subgroup of the Lorentz group whose transformations leave the
four-momentum of a given particle invariant.  In the present case,
the little group consists of transformations on a given density matrix
which will leave that matrix invariant.  It is shown in this paper
that the little group for pure states is like that for massless
particles, while the little group for impure states is like that for
massive particles.  The transition of the little group from a pure
to impure state is discussed in detail.

In Sec.~\ref{formul}, we show how each element in the two-beam
interferometer system corresponds to a transformation matrix in the
Lorentz group.  The combined effect is the two-by-two representation
of the six-parameter Lorentz group.
In Sec.~\ref{denma}, it is pointed out that the coherency matrix
can also be defined for the interferometer system and that this matrix
serves as the density matrix.  The transformation property of the
density matrix is discussed in detail.
In Sec.~\ref{little}, we introduce the little group which will leave
a given density matrix invariant.  It is noted that the little group
for pure states has a symmetry property quite different from that
for impure states.
In Sec.~\ref{decohm}, the decoherence matrices are discussed in
detail.  Although the augmentation of this matrix to the Lorentz group
leads to a large group, there exist subgroups exhibiting symmetry
properties familiar to us.  Possible experiments with the decoherence
matrix are suggested.

\section{Formulation of the Problem}\label{formul}
Typically, one beam is divided into two by a beam splitter.  We can
write the incoming beam as
\begin{equation}\label{expo1}
\Psi = \pmatrix{\psi_{1} \cr \psi_{2}} =
\pmatrix{ \exp{\left\{i(kz - \omega t)\right\}} \cr 0} .
\end{equation}
Then, the beam splitter can be written in the form of a rotation
matrix~\cite{sand99}:
\begin{equation}\label{rot22}
R(\theta) = \pmatrix{\cos(\theta/2) & -\sin(\theta/2) \cr
\sin(\theta/2) & \cos(\theta/2) } ,
\end{equation}
which transforms the column vector of Eq.(\ref{expo1}) into
\begin{equation}\label{expo2}
\pmatrix{\psi_{1} \cr \psi_{2}} =
\pmatrix{[\cos(\theta/2)]\exp{\left\{i(kz - \omega t)\right\}} \cr
-[\sin(\theta/2)\exp{\left\{i(kz - \omega t)\right\}} } .
\end{equation}
The first beam $\psi_{1}$ of Eq.(\ref{expo1}) is now split into
$\psi_{1}$ and $\psi_{2}$ of Eq.(\ref{expo2}).  The intensity is
conserved.  If the rotation angle $\theta$ is -$\pi/4$, the
initial beam is divided into two beams of the same intensity and
the same phase~\cite{campos89}.

These two beams go through two different optical path lengths,
resulting in a phase difference.  If the phase difference is
$\phi$, the phase shift matrix is
\begin{equation}\label{shif22}
P(\phi) = \pmatrix{e^{-i\phi/2} & 0 \cr 0 & e^{i\phi/2}} .
\end{equation}

When reflected from mirrors, or while going through beam splitters,
there are intensity losses for both beams.  The rate of loss is not
the same for the beams.  This results in the attenuation matrix
of the form
\begin{equation}\label{atten}
\pmatrix{e^{-\eta_{1}} & 0 \cr 0 & e^{-\eta_{2}}} =
e^{-(\eta_{1} + \eta_{2})/2} \pmatrix{e^{\eta/2} & 0 \cr 0 &
e^{-\eta/2}}
\end{equation}
with $\eta = \eta_{2} - \eta_{1}$ .
This attenuator matrix tells us that the electric fields are
attenuated at two different rates.  The exponential factor
$e^{-(\eta_{1} + \eta_{2})/2}$ reduces both components at the same
rate and does not affect the degree of polarization.  The effect of
polarization is solely determined by the squeeze matrix
\begin{equation}\label{sq22}
S(\eta) = \pmatrix{e^{\eta/2} & 0 \cr 0 & e^{-\eta/2}} .
\end{equation}

In the detector or the beam synthesizer, the two beams undergo a
superposition.  This can be achieved by the rotation matrix like the
one given in Eq.(\ref{rot22})~\cite{sand99}.   For instance, if the
angle $\theta$ is $90^{o}$, the rotation matrix takes the form
\begin{equation}
{1 \over \sqrt{2}}\pmatrix{1 & -1 \cr 1 & 1} .
\end{equation}
If this matrix is applied to the column vector of Eq.(\ref{expo2}),
the result is
\begin{equation}
{1 \over \sqrt{2}} \pmatrix{\psi_{1} - \psi_{2} \cr
\psi_{1} + \psi_{2}} .
\end{equation}
The upper and lower components show the interferences with negative
and positive signs respectively.

We have shown in our previous papers that repeated applications of the
rotation matrices of the form of Eq.(\ref{rot22}), shift matrices of
the form of Eq.(\ref{shif22}) and squeeze matrices of the form of
Eq.(\ref{sq22}) lead to a two-by-two representation of the six-parameter
Lorentz group.  The transformation matrix in general takes the form
\begin{equation}\label{lt22}
G = \pmatrix{\alpha & \beta \cr \gamma & \delta} ,
\end{equation}
applicable to the column vector of Eq.(\ref{expo1}), where all four
elements are complex numbers with the condition that the determinant
of the matrix be one.

Although we can borrow all the elegant mathematical identities of the
two-by-two representations of the Lorentz group, this formalism does
not allow us to describe the loss of coherence within the interferometer
system.  In order to study this effect, we have to construct the
coherency matrix:
\begin{equation}\label{cocy}
C = \pmatrix{S_{11} & S_{12} \cr S_{21} & S_{22}},
\end{equation}
with
\begin{eqnarray}\label{sii}
&{}& S_{11} = <\psi_{1}^{*}\psi_{1}>  , \qquad
S_{22} = <\psi_{2}^{*}\psi_{2}> , \nonumber \\[2ex]
&{}& S_{12} = <\psi_{1}^{*}\psi_{2}> ,  \qquad
S_{21} = <\psi_{2}^{*}\psi_{1}> .
\end{eqnarray}
It is sometimes more convenient to use the following combinations of
parameters.
\begin{eqnarray}\label{stokes}
&{}& S_{0} = S_{11} + S_{22}, \cr
&{}& S_{1} = S_{11} - S_{22}, \cr
&{}& S_{2} = S_{12} + S_{21}, \cr
&{}& S_{3} = -i\left(S_{12} - S_{21}\right).
\end{eqnarray}
These four parameters are called the Stokes parameters in the
literature~\cite{shur62,hecht70}, usually in connection with polarized
light waves.  In the present paper, we are applying these parameters
to two separate beams in a given interferometer system.

The Stokes parameters, originally developed for polarization optics,
are becoming applicable to other branches of physics dealing with
two orthogonal states.  In this paper, we are using these parameters
for interferometers.

We have shown previously~\cite{hkn97} that the four-by-four
transformation matrices applicable to the Stokes parameters are
like Lorentz-transformation matrices applicable to the space-time
Minkowskian vector $(t, z, x, y)$.
This allows us to study space-time symmetries in terms of the Stokes
parameters which are applicable to interferometers.
Let us first see how the rotation matrix of Eq.(\ref{rot22}) is
translated into the four-by-four formalism.  In this case,
\begin{equation}
\alpha = \delta = \cos(\theta/2), \qquad
\gamma = -\beta = \sin(\theta/2) .
\end{equation}
The corresponding four-by-four matrix takes the form~\cite{hkn99}
\begin{equation}\label{rot44}
R(\theta) = \pmatrix{1 & 0 & 0 & 0 \cr
0 & \cos\theta & -\sin\theta & 0  \cr
0 & \sin\theta & \cos\theta & 0 \cr
0 & 0 & 0 & 1} .
\end{equation}

Let us next see how the phase-shift matrix of Eq.(\ref{shif22}) is
translated into this four-dimensional space.  For this two-by-two
matrix,
\begin{equation}
\alpha = e^{-i\phi/2} , \qquad \beta = \gamma = 0 , \qquad
\delta = e^{i\phi/2} .
\end{equation}
For these values, the four-by-four transformation matrix
takes the form~\cite{hkn99}
\begin{equation}\label{shif44}
P(\phi) = \pmatrix{1 & 0 & 0 & 0 \cr 0 & 1 & 0 & 0  \cr
0 & 0 & \cos\phi & -\sin\phi \cr 0 & 0 & \sin\phi & \cos\phi} .
\end{equation}
For the squeeze matrix of Eq.(\ref{sq22}),
\begin{equation}
\alpha = e^{\eta/2}, \qquad \beta = \gamma = 0 , \qquad
\delta = e^{-\eta/2} .
\end{equation}
As a consequence, its four-by-four equivalent is
\begin{equation}\label{sq44}
S(\eta) = \pmatrix{\cosh\eta & \sinh\eta & 0 & 0 \cr
\sinh\eta & \cosh\eta & 0 & 0 \cr
0 & 0 & 1 & 0 \cr 0 & 0 & 0 & 1} .
\end{equation}
If the above matrices are applied to the four-dimensional
Minkowskian space of $(t, z, x, y)$, the above squeeze matrix will
perform a Lorentz boost along the $z$ or $S_{1}$ axis with $S_{0}$ as
the time variable.  The rotation matrix of Eq.(\ref{rot44}) will
perform a rotation around the $y$ or $S_{3}$ axis, while the phase
shifter of Eq.(\ref{shif44}) performs a rotation around the $z$ or
the $S_{1}$ axis.  Matrix multiplications with $R(\theta)$ and
$P(\phi)$ lead to the three-parameter group of rotation matrices
applicable to the three-dimensional space of $(S_{1}, S_{2}, S_{3})$.

The phase shifter  $P(\phi)$ of Eq.(\ref{shif44}) commutes with the
squeeze matrix of Eq.(\ref{sq44}), but the rotation matrix $R(\theta)$
does not.  This aspect of matrix algebra leads to many interesting
mathematical identifies which can be tested in laboratories.  One of
the interesting cases is that we can produce a rotation by performing
three squeezes~\cite{hkn99}.
Another interesting case is a combination of squeeze and rotation
will produce a matrix which will convert numerical multiplication
into addition.  This aspect known as the Iwasawa decomposition is
discussed in detail in Ref.~\cite{hkn99}.

\section{Density Matrices and Their Little Groups}\label{denma}

According to the definition of the density matrix~\cite{fey72}, the
coherency matrix of Eq.(\ref{cocy}) is also the density matrix.
Since we discussed transformation properties of coherency matrices
in our earlier papers~\cite{hkn97josa,hkn97}, we can start here with
those results on this subject.

The most effective way of formulating the symmetry property of a given
physical system is to construct a group of transformations which
leave the system invariant.  This concept was originally developed
by Wigner~\cite{wig39} for internal space-time symmetries of
relativistic particles.  Wigner's little group is the maximal subgroup
of the Lorentz group whose transformations leave the four-momentum of
a given particle invariant.  For instance, for a particle at rest,
the little group is the three-parameter rotation group.  The rotations
do not change the four-momentum of the particle, even though they
change the direction of the spin.
There are also massless particles which cannot be brought to rest.
This is the reason why the little group for a massive particle is
different from that of the massless particle.  The little group for
massless particles is like (or locally isomorphic to) the
two-dimensional Euclidean group~\cite{wig39,knp86}.

Indeed, in Ref.\cite{hkn99}, we discussed Wigner rotations and Iwasawa
decompositions rotations applicable to massive and massless particles
respectively and how these little-group transformations can be applied
to the Stokes four-vectors.  In this section, we shall see that the
Stokes vectors for pure and impure states are like the four-momentum of
the massless and massive particles respectively.

In the following discussion, we will need transformations of the Stokes
four-vectors and the corresponding transformations of the two-by-two
density matrices.  We are quite familiar with four-by-four matrices
applicable to the Stokes vectors.  For the two-by-two density matrices,
the transformation takes the form

Under the influence of the $G$ transformation given in Eq.(\ref{lt22}),
this coherency matrix is transformed as
\begin{eqnarray}\label{trans22}
&{}& C' = G\,C\,G^{\dagger} =
\pmatrix{S'_{11} & S'_{12} \cr S'_{21} & S'_{22}}  \nonumber \\[2ex]
&{}&\hspace{5ex} = \pmatrix{\alpha & \beta \cr \gamma & \delta}
\pmatrix{S_{11} & S_{12} \cr S_{21} & S_{22}}
\pmatrix{\alpha^{*} & \gamma^{*} \cr \beta^{*} & \delta^{*}} ,
\end{eqnarray}
where $C$ and $G$ are the density matrix and the transformation matrix
given in Eq.(\ref{cocy}) and Eq.(\ref{lt22}) respectively.  According
to the basic property of the Lorentz group, these transformations do
not change the determinant of the density matrix $C$.  Transformations
which do not change the determinant are called unimodular
transformations.

As we shall see in this section, the determinant for pure states is
zero, while for that for mixed states does not vanish.  Is there then
a transformation matrix which will change this determinant within the
Lorentz group.  The answer is No.  This is the basic issue we would
like to address in this section.

If the phase difference between the two waves remains intact, the the
system is said to in a pure state, and the density matrix can be
brought to the form
\begin{equation}\label{pure22}
\pmatrix{1 & 0 \cr 0 & 0} ,
\end{equation}
through the transformation of Eq.(\ref{trans22}) with a suitable
choice of the $G$ matrix.  For the pure state, the Stokes four-vector
takes the form
\begin{equation}\label{pure4}
\pmatrix{1 \cr 1 \cr 0 \cr 0} .
\end{equation}

In order to study the symmetry properties of the density matrix, let
us ask the following question.  Is there a group of transformation
matrices which will leave the above density matrix invariant?
In answering this question, it is more convenient to use the Stokes
four-vector.  The column vector of Eq.(\ref{pure4}) is invariant under
the operation of the phase shifter $P(\phi)$ of Eq.(\ref{shif44}).
In addition, it is invariant under the following two matrices:
\begin{eqnarray}\label{d1d2}
&{}& F_{1}(u) = \pmatrix{ 1 + u^{2}/2  & - u^{2}/2 & u  & 0  \cr
   u^{2}/2  & 1 - u^{2}/2 & u & 0   \cr
   u  & -u & 1 & 0 \cr
   0 & 0 & 0 & 1 } ,  \nonumber \\[2ex]
&{}& F_{2}(v) = \pmatrix{ 1 + v^{2}/2  & - v^{2}/2 & 0  & v  \cr
   v^{2}/2  & 1 - v^{2}/2 & 0 & v   \cr
   0  & 0 & 1 & 0 \cr
   u & -v & 0 & 1 } .
\end{eqnarray}
These mathematical expressions were first discovered by Wigner in
1939~\cite{wig39} in connection with the internal space-time
symmetries of relativistic particles.  They went through a stormy
history, but it is gratifying to note that they serve a useful purpose
for studying interferometers where each matrix corresponds to
an operation which can be performed in laboratories.

The $F_{1}$ and $F_{2}$ matrices commute with each other, and the
multiplication of these leads to the form
\widetext
\begin{equation}\label{d44}
F_{2}(u)F_{2}(v)
= \pmatrix{1 + (u^{2} + v^{2})/2  & - (u^{2} + v^{2})/2 & u & u \cr
   (u^{2} + v^{2})/2  & 1 - (u^{2} + v^{2})/2 & u & v   \cr
   u  & -u & 1 & 0 \cr
   v & -v & 0 & 1 } .
\end{equation}
\narrowtext
\noindent This matrix contains two parameters.

Let us go back to the phase-shift matrix of Eq.(\ref{shif44}).
This matrix also leaves the Stokes vector of Eq.(\ref{pure4})
invariant.  If we define the ``little group'' as the maximal subgroup
of the Lorentz group which leaves a Stokes vector invariant, the
little group for the Stokes vector of Eq.(\ref{pure4}) consists of the
transformation matrices given in Eq.(\ref{shif44}) and Eq.(\ref{d44}).

Next, if the phase relation is completely random, and the first and
second components have the same amplitude, the density matrix becomes
\begin{equation}\label{imp22}
\pmatrix{1/2 & 0 \cr 0 & 1/2} .
\end{equation}
Here is the question: Is there a two-by-two matrix which will
transform the pure-state density matrix of Eq.(\ref{pure22}) into the
impure-state matrix of Eq.(\ref{imp22})?  The answer within the system
of matrices of the form given in Eq.(\ref{lt22}) is No, because the
determinant of the pure-state density matrix is zero while that for
the impure-state matrix is $1/4$.  Is there a way to deal with this
problem?  We shall return to this problem in Sec.~\ref{decohm}.
In this section, we restrict ourselves to the unimodular transformation
of Eq.(\ref{trans22}) which preserves the value of the determinant of
the density matrix.  The Stokes four-vector corresponding to the above
density matrix is
\begin{equation}\label{imp4}
\pmatrix{1 \cr 0 \cr 0 \cr 0} .
\end{equation}
This vector is invariant under both the rotation matrix of
Eq.(\ref{rot44}) and the phase shift matrix of Eq.(\ref{shif44}).
Repeated applications of these matrices lead to a three-parameter
group of rotations applicable to the three-dimensional space of
$(S_{1}, S_{2}, S_{3})$.

Not all the impure-state density matrices take the form of
Eq.(\ref{imp22}).  In general, if they are brought to a diagonal
form, the matrix takes the form
\begin{equation}\label{impp22}
{1 \over 2}\pmatrix{1 + \cos\chi & 0 \cr 0 & 1 - \cos\chi} ,
\end{equation}
and the corresponding Stokes four-vector is
\begin{equation}\label{impp4}
e^{-\eta} \pmatrix{\cosh\eta \cr \sinh\eta \cr 0 \cr 0} ,
\end{equation}
with
\begin{equation}
\eta = {1 \over 2}\ln{1 + \cos\chi \over 1 - \cos\chi} .
\end{equation}
The matrix which transforms Eq.(\ref{imp4}) to Eq.(\ref{impp4}) is
the squeeze matrix of Eq.(\ref{sq44}).
The question then is whether it is possible to transform the pure
state of Eq.(\ref{pure4}) to the impure state of Eq.(\ref{impp4}) or
to Eq.(\ref{imp4}).

In order to see the problem in terms of the two-by-two density matrix,
let us go back to the pure-state density matrix of Eq.(\ref{pure22}).
Under the rotation of Eq.(\ref{rot22}),
\begin{eqnarray}
&{}& \pmatrix{\cos(\chi/2) & -\sin(\chi/2) \cr \sin(\chi/2) &
\cos(\chi/2) }  \pmatrix{1 & 0 \cr 0 & 0} \nonumber\\[2ex]
&{}& \times \pmatrix{\cos(\chi/2) &  \sin(\chi/2) \cr
-\sin(\chi/2) & \cos(\chi/2) } ,
\end{eqnarray}
the pure-state density matrix becomes
\begin{equation}
{1 \over 2} \pmatrix{1 + \cos\chi & \sin\chi \cr
                      \sin\chi & 1 - \cos\chi} .
\end{equation}

For the present case of two-by-two density matrices, the trace of
the matrix is one for both pure and impure cases.  The trace of
the $(matrix)^{2}$ is one for the pure state, while it is less
than one for impure states.

The next question is whether there is a two-by-two matrix which will
eliminate the off-diagonal elements of the above expression that
will also lead to the expression of Eq.(\ref{impp22}).
In order to answer this question, let us note that the determinant of
the density matrix vanishes for the pure state, while it is non-zero
for impure states.  The Lorentz-like transformations of
Eq.(\ref{trans22}) leave the determinant invariant.  Thus, it is not
possible to transform a pure state into an impure state by means of
the transformations from the six-parameter Lorentz group.  Then
is it possible to achieve this purpose using two-by-two matrices
not belonging to this group.  We do not know the answer to this
question.  We are thus forced to resort to four-by-four matrices
applicable to the Stokes four-vector.

\section{Decoherence Effects on the Little Groups}\label{little}
We are interested in a transformation which will change the density
matrix of Eq.(\ref{pure22}) to Eq.(\ref{imp22}).  For this purpose,
we can use the Stokes four-vector consisting of the four elements of
the density matrix.  The question then is
whether it is possible to find a transformation matrix which will
transform the pure-state four-vector of Eq.(\ref{pure4}) to the
impure-state four-vector of Eq.(\ref{imp4}).

Mathematically, it is more convenient to ask whether the inverse of
this process is possible: whether it is possible to transform the
four-vector of Eq.(\ref{imp4}) to that of Eq.(\ref{pure4}).  This is
known in mathematics as the contraction of the three-dimensional
rotation group into the two-dimensional Euclidean group~\cite{knp86}.
Let us apply the squeeze matrix of Eq.(\ref{sq44}) to the four-vector
of Eq.(\ref{imp4}).  This can be written as
\begin{equation}\label{sqimp}
\pmatrix{\cosh\eta & \sinh\eta & 0 & 0 \cr
\sinh\eta & \cosh\eta & 0 & 0 \cr
0 & 0 & 1 & 0 \cr 0 & 0 & 0 & 1}
\pmatrix{1 \cr 0 \cr 0 \cr 0} =
\pmatrix{\cosh\eta \cr \sinh\eta \cr 0 \cr 0} .
\end{equation}
After an appropriate normalization, the right-hand side of the above
equation becomes like the pure-state vector of Eq.(\ref{pure4}) in the
limit of large $\eta$, as $\cosh\eta$ becomes equal to $\sinh\eta$
in the infinite-$\eta$ limit.  This transformation is from a mixed
state to a pure or almost-pure state.  Since we are interested in the
transformation from the pure state of Eq.(\ref{pure4})  to the impure
state of Eq.(\ref{imp4}), we have to consider an inverse of the above
equation:
\begin{equation}\label{inver1}
\pmatrix{\cosh\eta & -\sinh\eta & 0 & 0 \cr
-\sinh\eta & \cosh\eta & 0 & 0 \cr 0 & 0 & 1 & 0 \cr 0 & 0 & 0 & 1}
\pmatrix{\cosh\eta \cr \sinh\eta \cr 0 \cr 0}  =
\pmatrix{1 \cr 0 \cr 0 \cr 0} .
\end{equation}
However, the above equation does not start with the pure-state
four-vector.  If we apply the same matrix to the pure state matrix,
the result is
\begin{equation}
\pmatrix{\cosh\eta & -\sinh\eta & 0 & 0 \cr
-\sinh\eta & \cosh\eta & 0 & 0 \cr 0 & 0 & 1 & 0 \cr 0 & 0 & 0 & 1}
\pmatrix{1 \cr 1 \cr 0 \cr 0}
= e^{-\eta} \pmatrix{1 \cr 1 \cr 0 \cr 0} .
\end{equation}
The resulting four-vector is proportional to the pure-state four-vector
and is definitely not an impure-state four-vector.

The inverse of the transformation of Eq.(\ref{sqimp}) is not capable
of bringing the pure-state vector into an impure-state vector.  Let us
go back to Eq.(\ref{sqimp}), it is possible to bring a impure-state
into a pure state only in the limit of infinite $\eta$.  Otherwise,
it is not possible.  It is definitely not possible if we take into
account experimental considerations.

The story is different for the little groups.  Let us start with
the rotation matrix of Eq.(\ref{rot44}), and apply to this matrix
the transformation matrix of Eq.(\ref{sqimp}).  Then
\widetext
\begin{equation}\label{3mats}
\pmatrix{\cosh\eta & \sinh\eta & 0 & 0 \cr
\sinh\eta & \cosh\eta & 0 & 0 \cr
0 & 0 & 1 & 0 \cr 0 & 0 & 0 & 1}
\pmatrix{1 & 0 & 0 & 0 \cr 0 & \cos\theta & -\sin\theta & 0 \cr
0 & \sin\theta & \cos\theta & 0 \cr 0 & 0 & 0 & 1}
\pmatrix{\cosh\eta & -\sinh\eta & 0 & 0 \cr
-\sinh\eta & \cosh\eta & 0 & 0 \cr
0 & 0 & 1 & 0 \cr 0 & 0 & 0 & 1} .
\end{equation}
\narrowtext
\noindent If $\eta$ is zero, the above expression becomes the rotation
matrix of Eq.(\ref{rot44}).  If $\eta$ becomes infinite, it becomes
the little-group matrix $F_{1}(u)$ of Eq.(\ref{d1d2}) applicable to
the pure state of Eq.(\ref{pure4}).  The details of this calculation
for the case of Lorentz transformations are given in the 1986 paper
by Han {\it et al.}~\cite{hks86jm}.
We are then led to the question of whether one
little-group transformation matrix can be transformed from the other.

If we carry out the matrix algebra of Eq.(\ref{3mats}), the result is
\widetext
\begin{equation}
\pmatrix{1 + \alpha u^{2} w/2 & -\alpha u^{2} w/2 & \alpha uw & 0 \cr
\alpha u^{2} w/2 & 1 - u^{2} w/2 & uw & 0 \cr
\alpha uw & -uw & 1 - (1 - \alpha^{2}) u^{2} w/2 & 0 \cr
0 & 0 & 0 & 1} ,
\end{equation}
where
\begin{equation}\label{anal}
\alpha = \tanh\eta , \qquad
u = - 2\,\tan\left({\theta \over 2}\right), \qquad
w = { 1 \over 1 + (1 - \alpha^{2})\tan^{2}(\theta/2)}.
\end{equation}
\narrowtext
\noindent  If $\alpha = 0$, the above expression becomes the rotation
matrix of Eq.(\ref{rot44}).  If $\alpha = 1$, it becomes the $F_{1}$
matrix of Eq.(\ref{d1d2}).  Here we used the parameter $\alpha$
instead of $\eta$.  In terms of this parameter, it is possible to make
an analytic continuation from the pure state with $\alpha = 1$ to an
impure state with $\alpha < 1$ including $\alpha = 0$.

On the other hand, we should keep in mind that the determinant of the
density matrix is zero for the pure state, while it is non-zero for
all impure states.  For $\alpha = 1$, the determinant vanishes, but
it is nonzero and stays the same for all non-zero values of $\alpha$
less than one and greater than or equal to zero.  The analytic
expression of Eq.(\ref{anal}) hides this singular nature of the
little group~\cite{hks86jm}.

\section{Decoherence Matrices}\label{decohm}
We are interested in the decoherence effect on the density matrix.
We are particularly interested in the mechanism where the off-diagonal
elements $S_{12}$ and $S_{21}$ become smaller due to time average or
phase-randomizing process~\cite{raymer97}.  If this happens, we can
apply to the Stokes four-vector the following decoherence matrix.
\begin{equation}
\pmatrix{1 & 0 & 0 & 0 \cr 0 & 1 & 0 & 0 \cr
0 & 0 & e^{-2\lambda} & 0 \cr 0 & 0 & 0 & e^{-2\lambda}} ,
\end{equation}
which can also be written as
\begin{equation}\label{decoh1}
e^{-\lambda} \pmatrix{e^{\lambda} & 0 & 0 & 0 \cr
0 & e^{\lambda} & 0 & 0 \cr 0 & 0 & e^{-\lambda} & 0 \cr
0 & 0 & 0 & e^{-\lambda}} ,
\end{equation}
where $e^{-\lambda}$ is the overall decoherence factor.  For
convenience, we define the decoherence matrix as
\begin{equation}\label{dlam}
D(\lambda) = \pmatrix{e^{\lambda} & 0 & 0 & 0 \cr
0 & e^{\lambda} & 0 & 0 \cr 0 & 0 & e^{-\lambda} & 0 \cr
0 & 0 & 0 & e^{-\lambda}} .
\end{equation}
This matrix cannot be constructed from the six-parameter Lorentz
group applicable to the Stokes four-vectors.

If we combine this decoherence matrix with the Lorentz group, the
result will be a fifteen-parameter group of four-by-four matrices
isomorphic to $O(3,3)$ which is beyond the scope of the present
paper~\cite{hkn95jm}.  In order to extract the symmetry of physical
interest, let us go back to the four-by-four matrices
$R(\theta), P(\phi)$, and $S(\eta)$ of Eq.(\ref{rot44}),
Eq.(\ref{shif44}), and Eq.(\ref{sq44}) respectively.  The phase-shift
matrix of Eq.(\ref{shif44}) commutes with the decoherence matrix.

As we discussed in our earlier paper on polarization
optics~\cite{hkn97}, the decoherence matrix and the rotation matrix
will lead to two-dimensional squeeze transformations applicable to
the two-component vector
\begin{equation}\label{va}
V_{A} = \pmatrix{S_{1} \cr S_{2}} .
\end{equation}
The four-by-four $D(\lambda)$ matrix of Eq.(\ref{dlam}) and the
rotation matrix of Eq.(\ref{rot44}) become reduced to
\begin{eqnarray}\label{dara}
&{}& D_{A}(\lambda) = \pmatrix{e^{\lambda} & 0 \cr
0 & e^{-\lambda}} , \nonumber \\[2ex]
&{}& R_{A}(\theta) = \pmatrix{\cos\theta & -\sin\theta \cr
\sin\theta & \cos\theta} .
\end{eqnarray}
As for the remaining components of the Stokes parameters, we can
define another two-component vector as
\begin{equation}
V_{B} = \pmatrix{S_{0} \cr S_{3}} .
\end{equation}
The decoherence matrix applicable to this two-component vector is
\begin{equation}
D_{B}(\lambda) = \pmatrix{e^{\lambda} & 0 \cr 0 & e^{-\lambda}} ,
\end{equation}
but the rotation matrix does not change the two-component vector
$V_{B}$.

Let us go back to the two-dimensional space of $V_{A}$, and its
two-by-two transformation matrices.  The matrices $D_{A}(\lambda)$
and $R_{A}(\theta)$ of Eq.(\ref{dara}) applicable are strikingly
similar to the two-by-two matrices given in Eq.(\ref{sq22}) and
Eq.(\ref{rot22}) respectively.  If we replace the parameters
$\eta$ in $S(\eta)$ and $\theta$ in $R(\theta)$ by $2\lambda$
and $2\theta$ respectively, they become $D_{A}$ and $R_{B}$ of
Eq.(\ref{dara}).

With these two matrices, we can repeat the calculations for the
Wigner rotations and Iwasawa decompostions discussed in our earlier
paper~\cite{hkn99}. It is possible to perform experiments to test these
mathematical relations.

\section*{Concluding Remarks}
In this paper, we have discussed two-beam interferometers within the
framework of the six-parameter Lorentz group.  It has been shown that
beam splitters and beam synthesizers can be represented by two-by-two
rotation matrices.  The phase shift can also be represented by
two-by-two rotation matrices applicable to spinor systems.  As for
attenuation, we introduced two-by-two squeeze matrices.  The combined
effect of these transformations leads to a two-by-two representation
of the six-parameter Lorentz group.

We have found that the mathematical formalism given in this paper is
identical to the formalism we presented in our earlier papers for
polarization optics.  In this series of papers, our purpose has been
to minimize the group theoretical language and write down formulas
close to what we observe in the real world.  In this paper, we were
able to by-pass completely the group theoretical formality known as
the Lie algebra of the Lorentz group consisting of generators and
their closed commutation relations.

With this improved mathematical technique, we discussed two-beam
physics in terms of the little groups using only matrices which are
realizable in laboratories.  It has been shown that the little groups
for pure and impure states are different.  It was noted that analytic
continuation from a pure state to an impure state is possible for
the little groups.  On the other hand, this transformation does
not exist within the six-parameter Lorentz group, but requires
an extra four-by-four matrix applicable to the Stokes four-vector,
called the decoherence matrix.

The augmentation of this decoherence matrix into the Lorentz group
will lead to a bigger group which is beyond the scope of this
paper~\cite{hkn95jm}.  However, this bigger group has $O(2,1)$-like
or $SU(1,1)$-like subgroups which are quite familiar to us from the
squeezed states of light, and the Lorentz group-formulation of the
polarization optics~\cite{hkn97}.  We are fortunate to observe,
within the framework of this decoherence matrix, mathematical
consequence which will lead to experiments on Wigner rotations and
Iwasawa decompositions which are possible in both polarization optics
and interferometers.

It will be a challenging problem to translate what we did in this
paper to the language of quantum optics.  The rotation
operations corresponding to phase shifts and rotations around the
direction of the propagation can be formulated in terms of the
two-mode squeezed states~\cite{ymk86}.  However, the squeeze
transformations discussed in this paper correspond to the loss
of intensity, which cannot be translated into quantum optics.  On the
other hand, the decoherence matrix can be accommodated into the
density-matrix formalism.  Indeed, they all are challenging problems.

Furthermore, unlike the case of polarization optics, there can be
more than two beams for interferometers.  For instance, three-beam
interferometer are quite common.  This will open up a new research
line for studying symmetry properties in optics.  The power of
group theoretical approaches is that we can establish the symmetry
properties in one branches of physics to those in a different field
using the isomorphism and/or homomorphism of group theory.  As for
the three-beam case, we are happy to note a recent paper by
Rowe {\it et al}.~\cite{rowe99}.

\end{document}